\documentclass[aps]{revtex4}
\usepackage{amsfonts}
\usepackage{amsmath}
\usepackage{amssymb,epsf}

\begin{document}

\title{Complexity of the Einstein-Born-Infeld-Massive Black holes}
\author{S. H. Hendi$^{1,2}$\footnote{
email address: hendi@shirazu.ac.ir} and B.
Bahrami-Asl$^{1}$\footnote{ email address:
banafsheh.bahrami@shirazu.ac.ir}} \affiliation{$^1$ Physics
Department and Biruni Observatory, College of Sciences, Shiraz
University, Shiraz 71454, Iran\\
$^2$ Research Institute for Astronomy and Astrophysics of Maragha
(RIAAM), Maragha, Iran}

\begin{abstract}
Motivated by interesting correspondence between computational
complexity in a CFT and the action evaluated on a WDW patch in the
bulk, we study the complexity of the Einstein-massive black holes
in the presence of BI nonlinear electrodynamic. The upper limit of
Llyod bound according to the WDW patch is investigated and it is
proved that Llyod bound is held.
\end{abstract}

\maketitle

\section{ Introduction}

Thermodynamical systems, obeying the laws of classical or quantum
mechanics, are characterized by a set of intensive and extensive
quantities. Among these quantities, the main two characteristics
of thermodynamical systems are entropy and temperature. It is
known that in the equilibrium state the former has the greatest
value and the later is a constant. However, one can use the first
law of thermodynamics to calculate the absolute temperature for
non-equilibrium situation. In order to understand the conception
of entropy, one has to focus on the concept of information which
is related to the measure of uncertainty in physical systems.
Considering an identical concept for the entropy and information
is a confusion and in order to prevent such misunderstanding, the
complexity statement is used to discuss characteristics of
physical thermodynamical systems. The value of the complexity of a
system can change due to some influences of environment or
strictly speaking, transmission of information between the
environment and system.

According to the Bekenstein suggested \cite{Bekenstein1973}, black
holes have entropy, and therefore, laws of thermodynamics can be
valid for them. Regarding black hole entropy, it is natural to
think about the existence of information inside the event horizon
as well as information paradox after Hawking radiation. In
addition, it is interesting to look for an approach for
distinguishing black holes with different information. The black
hole information is related to its entropy and consequently
complicatedness or complexity. As a result, measuring the
complexity of black holes may help us to resolve the information
paradox and quantum nature of black hole as well.

The formal definition of computational complexity in the context
of quantum mechanics is related to the minimum number of quantum
gates in quantum circuit which is required to prepare the boundary
state from a simple state \cite{Aspnes2017}. In other words,
complexity is the minimal difficulty of taking the system from a
simple reference state to a particular state of interest. In the
context of black hole physics, the complexity is the boundary
state complicatedness of a geometric property of the black hole
interior. It may suggest that the structure of black hole interior
is a geometric representation of a quantum circuit or is matched
with the geometry of tensor network, and therefore, one can use
the holography principle to estimate the complexity
\cite{susskindcc,susskindca,susskindhc,nielsenqc}.

In order to find a suitable relation for the complexity based on
AdS/CFT correspondence, we consider a situation of sending a
signal through Einstein-Rosen bridge (ERB). We need an appropriate
duality which connects the quantity of theoretical information
with a geometric concept for calculating the complicatedness. It
is known that the computational complexity of the boundary state
is proportional to the volume of ERB or in general case is
proportional to the volume of black hole interior
\cite{susskindcc}
\begin{equation}
C\sim \frac{V}{Gl},  \label{C-V}
\end{equation}
where $G$ is the Newton's constant and $l$ is a length scale which
is related to the AdS radius for large black holes and for small
ones it is proportional to the Schwarzschild radius
\cite{susskindcc,susskindhc}. Multiplying Eq. (\ref{C-V}) by $l$,
one can propose a new perspective of the complexity
\begin{equation}
C \sim \frac{W}{Gl^{2}},  \label{C-W}
\end{equation}
where $W=lV$ has the units of space-time volume and
$\frac{1}{l^{2}}$ is proportional to the cosmological constant of
the AdS space. Equation (\ref{C-W}) inspires new conjecture that
connects the complexity with the gravitational dynamics
\cite{susskindhc,susskindnc}
\begin{equation}
C\sim \frac{A}{\pi h}, \label{C-A}
\end{equation}
where $A$ is the action which is calculated by integrating of the
balk Lagrangian over $W$ with an appropriate boundary term.
Equation (\ref{C-A}) indicates complexity-action conjecture which
suggests that the complexity is proportional to the action. This
equation induces a deep connection between quantum information and
gravitational dynamics, and strictly speaking, it explains a
connection between tensor networks and geometry which means that
the geometry is defined by the smallest tensor network preparing
the state \cite{nielsenqc}.

Another interesting aspect of the complexity is its time
evolution. There is an interesting conjecture that bounds
variation of the complexity which is inspired by Llyod with the
following form \cite{llyod}
\begin{equation}
\frac{\partial }{\partial t}C\left( e^{-iHt}|\psi >\right) \leq \frac{%
2E_{\psi }}{\pi h}|\psi >,  \label{var-comp}
\end{equation}
where $E_{\psi }$ is the average energy of state $\psi$ which is
related to the ground state. For charged rotating black holes, Eq.
(\ref{var-comp}) reduces to \cite{susskindhc}
\begin{equation}
\frac{\partial }{\partial t}C\leq \frac{2}{\pi h}\left[ \left(
M-\mu Q-\Omega J\right) -\left( M-\mu Q-\Omega J\right)
_{ground\;\;state}\right], \label{bound}
\end{equation}
where $M$, $\mu $, $Q$, $\Omega $ and $J$ are, respectively, mass,
chemical potential, conserved charge, angular velocity and angular
momentum of the black hole.

For most cases, one may propose that the quantum complexity of
boundary state is equal to the classical action of spacetime in
the maximally extended black hole defined with respect to two
choices of time, on each boundary; the enclosed area is called the
Wheeler-De Witt (WDW) patch \cite{susskindca,susskindhc}.

As we mentioned, there are two complexity conjectures; the
complexity-volume duality and the complexity-action duality. The
complexity-volume conjecture states that the complexity of black
holes is dual to the volume of the black hole interior while the
complexity-action conjecture provides a relation between the
complexity of black holes and the action of the associated WDW
patch. In this paper, we are going to investigate the complexity
of black holes in Einstein-massive gravity in the presence of
nonlinear electrodynamics. Although Einstein theory is one of the
best theory with some correct predictions, there are some
mismatches that motivate one to generalize it. As an example we
refer the reader to the non-renormalizable properties of general
relativity which is arisen from the fact that this theory is
consistent with interaction of massless spin-$2$ gravitons. As a
result, it is logical to modifying general relativity to the case
of massive gravity with massive spin-$2$ particles. Massive
gravity has some advantages with respect to Einstein theory. Among
them, one may refer to explanation of accelerated expansion of the
universe without including dark energy and also renormalizable
property which helps us to understand the conceptions of quantum
gravity \cite
{fierzpauli,fierz,boulware,hassanprl,hassanhep,park,derham2010,derham2011,kurt}.
Different aspects of massive gravity have been investigated in
literature. AdS massive gravity is investigated in \cite{vegh2013}
\cite{hassan2011} while charged massive gravity is studied in
\cite{cai2015}. In addition, there are some interesting papers in
the context of massive gravity in the presence of nonlinear
electrodynamics
\cite{hendi1,hendi2,hendi3,hendi4,hendi5,hendi6,hendi7,hendi8,hendi9,hendi10}.
The main motivation of considering the nonlinear electrodynamics
is overcoming on the main problem of the Maxwell theory which is
the infinite self-energy of the point-like charges. In this
regard, Born and Infeld introduced a nonlinear electrodynamics
which is known as Born-Infeld (BI) theory \cite{born1934}.  One of
the interesting properties of the BI electrodynamics is that its
effective action arises in an open superstring theory and D-brains
with nonsingular self-energy of the point-like charges
\cite{fradkin1985,david1988,robert1989,gibbons2001} (we refer the
reader to see \cite{gibbons2003} for reviewing aspects of BI
theory in the context of string theory). Recently many papers have
published with the subject of complexity, which are investigated
the complexity of black holes in the presence of dilaton field,
Maxwell field and nonlinear electrodynamics with Einstein or
modified gravity theories
\cite{cai2017,swingle2017,an2018,cai2018,guo2017}. It seems that
this subject will be one of the hot topics for equipping the
classical theories of gravitation with some quantum
characteristics.

In this paper, we interested in studying complexity of the AdS
black holes in the massive gravity with BI electrodynamics. First,
we introduce the suitable action of the Einstein-BI-massive black
hole and its metric according to the symmetry of the spacetime.
Next, in the context of CA duality, the complexity of holographic
state dual to the Einstein-BI-massive black hole in the AdS space
is obtained and then growth of the complexity is calculated and
discussed.

\section{Einstein-Born-Infeld-massive Gravity}

Our starting point is the dRGT action for ghost-free massive
gravity with a nonlinear electrodynamics and negative cosmological
constant
\begin{equation}
S=-\frac{1}{16\pi }\int d^{d}x\sqrt{-g}\left[ R-2\Lambda
+L(\digamma )+m^{2}\sum\nolimits_{i=1}^{n}c_{i}U_{i}(g,f)\right]
\label{action}
\end{equation}%
in which $R$ is the scalar curvature of dynamical metric ($g_{\mu
\nu}$), $\Lambda =-\frac{(d-1)(d-2)}{2l^{2}}$ is the negative
cosmological constant and $L(\digamma )$ describes the Lagrangian
of a nonlinear model of electrodynamics which is called BI theory
\begin{equation}
L(\digamma )=4b ^{2}\left( 1-\sqrt{1+\frac{F_{\mu \nu }F^{\mu \nu }}{%
2 b ^{2}}}\right),  \label{born-infeld}
\end{equation}
where $ b $ is the Born-Infeld parameter and for $b \rightarrow
\infty $ the Born-Infeld theory reduces to Maxwell theory with
$L(\digamma )=-\digamma $, where $\digamma $ is the Maxwell
invariant $\digamma =F_{\mu \nu }F^{\mu \nu }$ with $F_{\mu \nu
}=\partial _{\mu }A_{\nu }-\partial _{\nu }A_{\mu }$ as the
Faraday tensor with the gauge potential $A_{\mu }$. The last term
of Eq. (\ref{action}) has a potential term role containing no
derivatives of the dynamical metric but it depends explicitly on a
non-dynamical symmetric reference metric $f_{\mu \nu}$. In
addition, $c_{i}$'s are some constants and $U_{i}$'s denote the
symmetric polynomials of the eigenvalues of $d\times d$ matrix
$\kappa _{\nu }^{\mu }=\sqrt{g^{\mu \alpha }f_{\alpha \nu }}$
which can be written as
\begin{eqnarray}
U_{1} &=&\left[ \kappa \right] ,  \nonumber \\
U_{2} &=&[\kappa ]^{2}-[\kappa ^{2}],  \nonumber \\
U_{3} &=&[\kappa ]^{3}-3[\kappa ][\kappa ^{2}]+2[\kappa ^{3}],  \nonumber \\
U_{4} &=&[\kappa ]^{4}-6[\kappa ^{2}][\kappa ]^{2}+8[\kappa
^{3}][\kappa
]+3[\kappa ^{2}]^{2}-6[\kappa ^{4}],  \nonumber \\
&&... \;.  \label{massive term}
\end{eqnarray}

It is worth mentioning that $U_{n}$'s have no contribution in the
field equations for $n \geq d$. Since higher order terms of
$U_{n}$'s ($4<n<d$) have no significant effect on the geometrical
behavior of the solutions, we restrict the solutions up to $U_4$.
Variation of the action with respect to the dynamical metric and
also gauge potential leads to the following field equations
\begin{equation}
G_{\mu \nu }+\Lambda g_{\mu \nu }-\frac{1}{2}g_{\mu \nu }L(\digamma )-\frac{%
2F_{\mu \lambda }F_{\nu }^{\lambda }}{\sqrt{1+\frac{F_{\mu \nu }F^{\mu \nu }%
}{2 b ^{2}}}}+m^{2}\chi _{\mu \nu }=0,  \label{FE1}
\end{equation}
\begin{equation}
\partial _{\mu }\left( \frac{\sqrt{-g}F^{\mu \nu }} {\sqrt{1+\frac{\digamma}{2 b ^{2}}}}\right) =0,  \label{FE2}
\end{equation}
where $G_{\mu \nu }$ is the Einstein tensor and $\chi _{\mu \nu }$
is related to the massive term with the following explicit form
\begin{eqnarray}
\chi _{\mu \nu } &=&-\frac{c_{1}}{2}\left( U_{1}g_{\mu \nu
}-\kappa _{\mu \nu }\right) -\frac{c_{2}}{2}\left( U_{2}g_{\mu \nu
}-2U_{1}\kappa _{\mu \nu }+2\kappa _{\mu \nu }^{2}\right)
-\frac{c_{3}}{2}(U_{3}g_{\mu \nu
}-3U_{2}\kappa _{\mu \nu }+  \nonumber \\
&&6U_{1}\kappa _{\mu \nu }^{2}-6\kappa _{\mu \nu }^{3})-\frac{c_{4}}{2}%
\left( U_{4}g_{\mu \nu }-4U_{3}\kappa _{\mu \nu }+12U_{2}\kappa
_{\mu \nu }^{2}-24U_{1}\kappa _{\mu \nu }^{3}+24\kappa _{\mu \nu
}^{4}\right)+...\; . \label{massive  contributions}
\end{eqnarray}

Now, we obtain static nonlinearly charged black holes in context
of massive gravity with adS asymptotes. For this purpose, we adopt
a static metric of $d$-dimensional spacetime in the following form
\begin{equation}
ds^{2}=-f(r)dt^{2}+\frac{1}{f(r)}dr^{2}+r^{2}h_{ij}dx^{i}dx^{j},\
\ \ i,j=1,2,3,...,d-2 ,  \label{metric}
\end{equation}
where $h_{ij}dx^{i}dx^{j}$ is a ($d-2$)-dimensional line element
for the Euclidian space with constant curvature $(d-2)(d-3)k$ and
volume $V_{d-2}$. We should note that the constant $k$ indicates
the boundary of $t=constant$ and $r=constant$, and it can be
negative, zero and positive which indicates hyperbolic, flat and
elliptic hypersurface, respectively. In addition, we consider the
following ansatz for the non-dynamic metric
\begin{equation}
f_{\mu \nu }=diag(0,0,c^{2}h_{ij}),  \label{euclidian}
\end{equation}
where $c$ is a positive constant. Using the mentioned ansatz for
$f_{\mu \nu }$, one can find that U$_{i}$'s are simplified as
\begin{equation}
U_{i}=\left(\frac{c}{r} \right)^{i}\prod\nolimits_{j=2}^{i}\left(
d-j\right). \label{ui}
\end{equation}

\begin{figure}[tbp]
$%
\begin{array}{cc}
\epsfxsize=6cm \epsffile{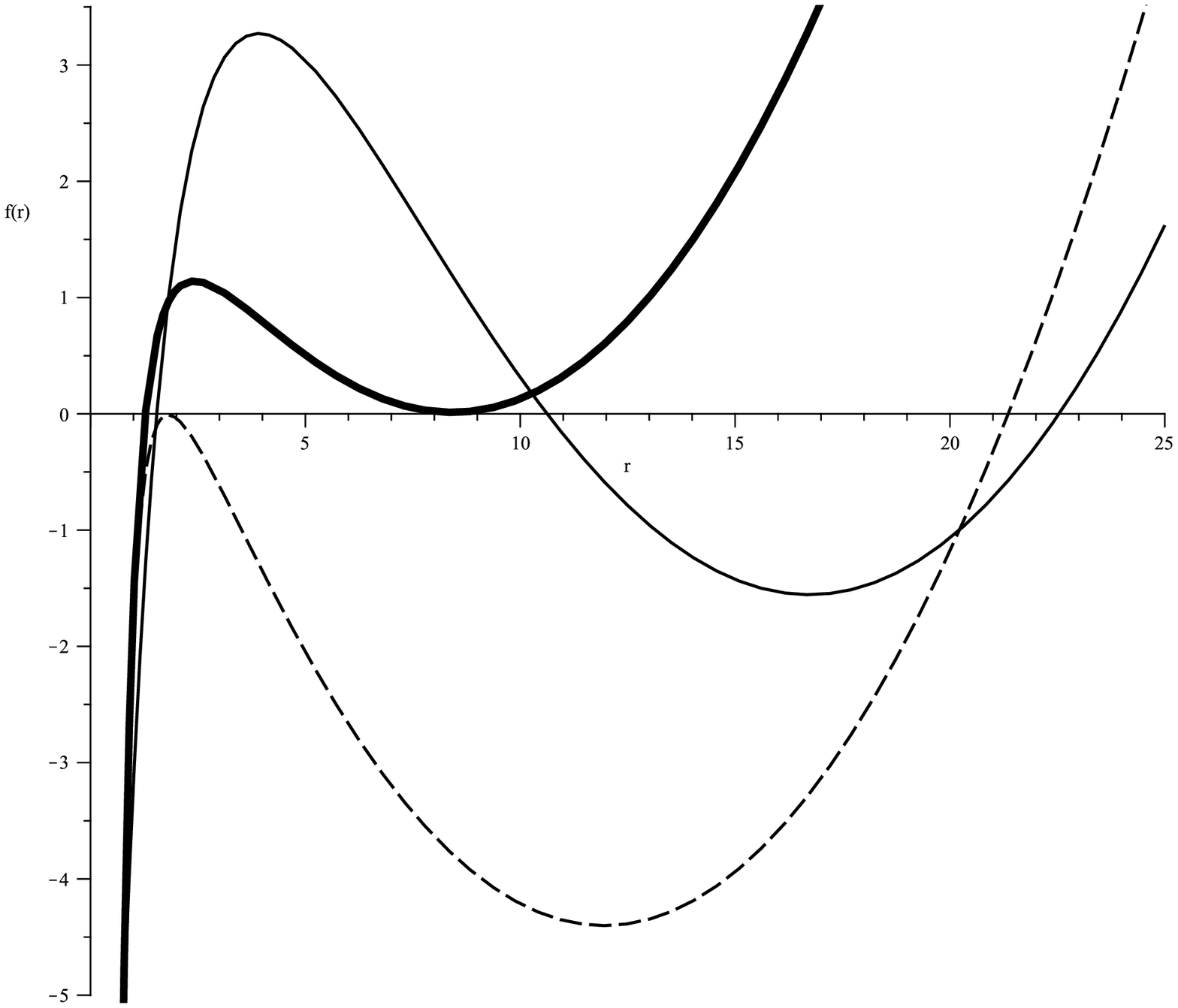} & \epsfxsize=5cm
\epsffile{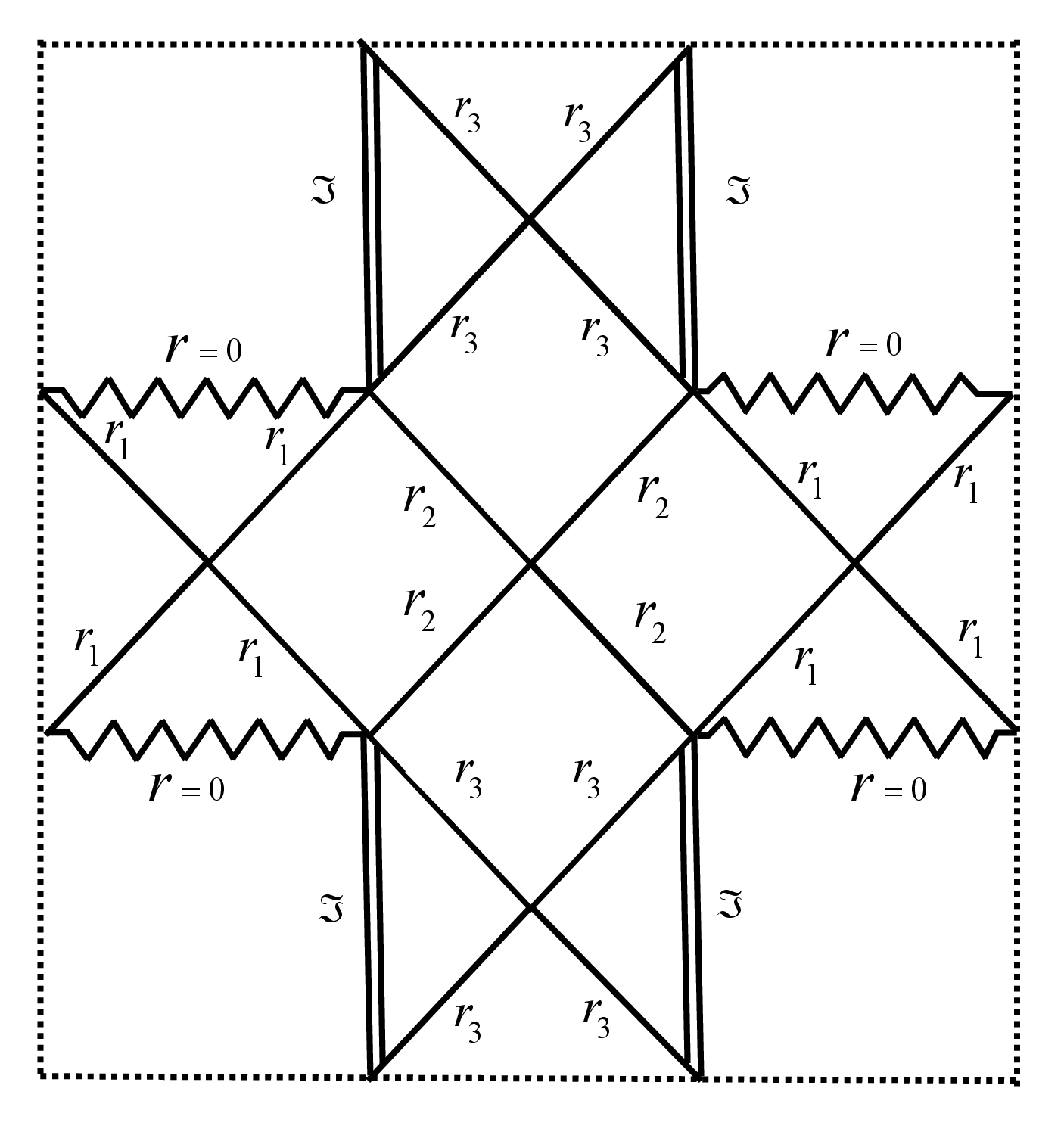}
\\
\epsfxsize=6cm \epsffile{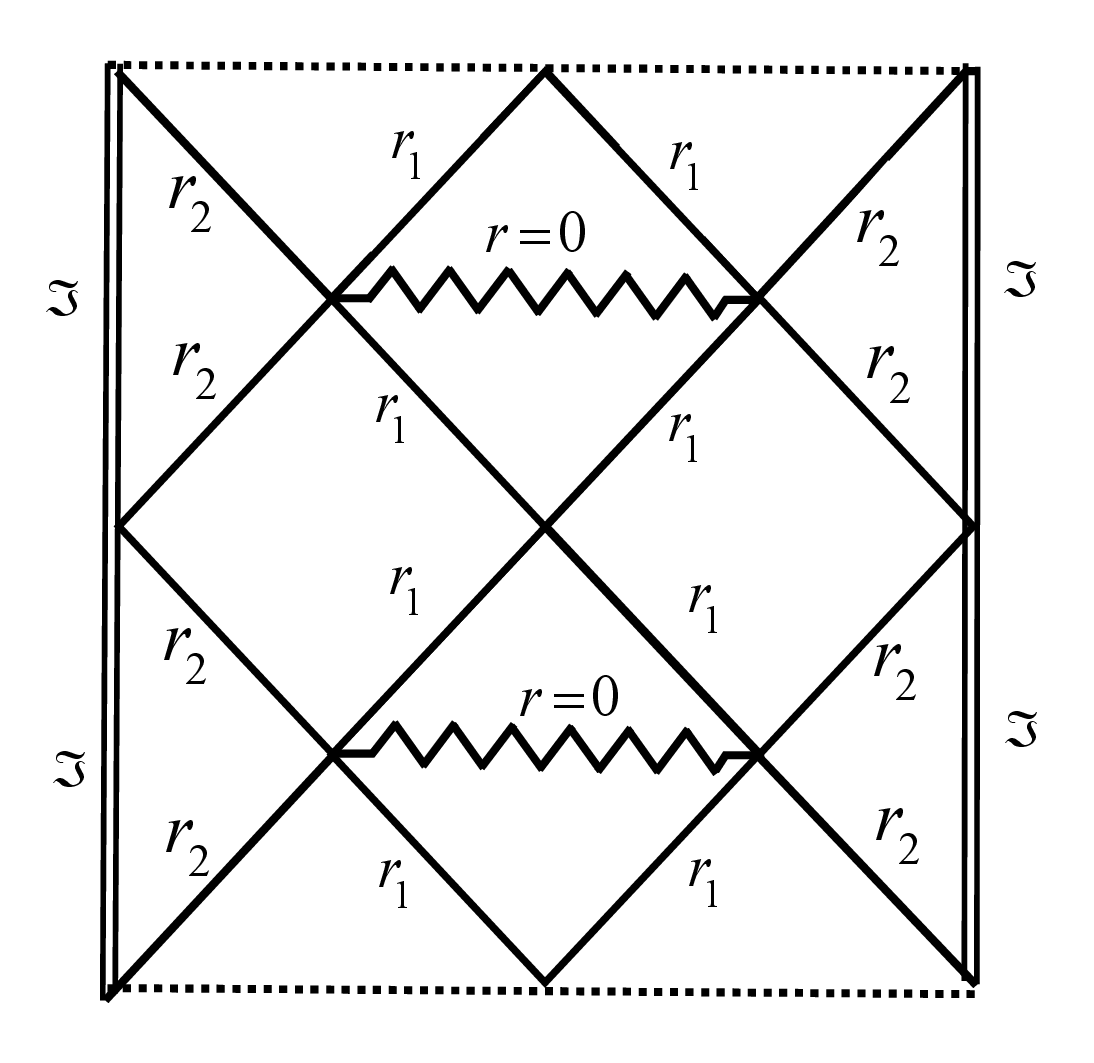} & \epsfxsize=5cm %
\epsffile{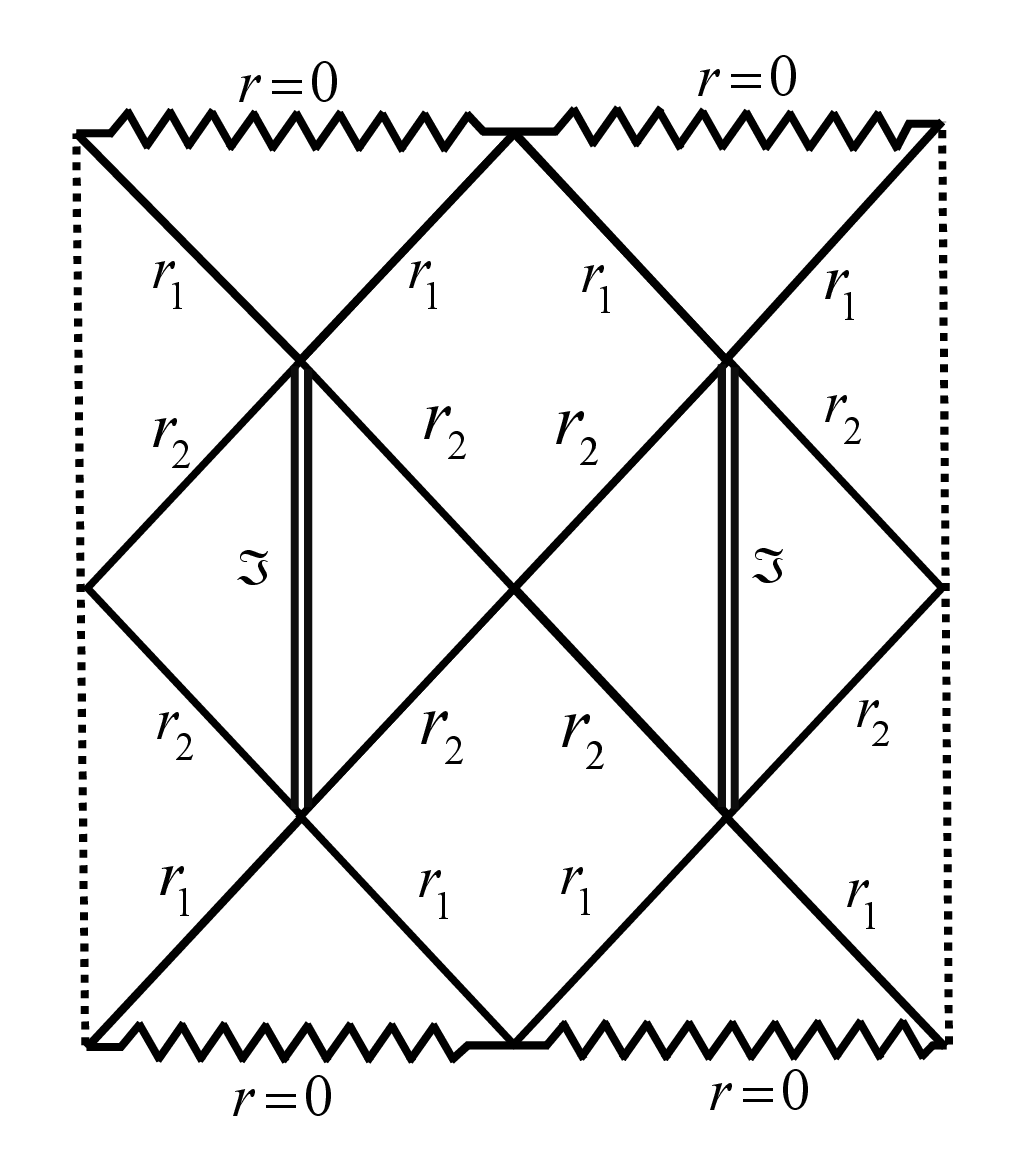}%
\end{array}
$%
\caption{The three cases for behaviors of the metric function and
their Penrose diagrams are plotted, Three horizons (continuous
line of metric function and related Penrose diagram in the
right-up panel), two horizons which inner one is extreme (dashed
line of metric function and related Penrose diagram in the
left-down panel) and two horizons which outer one is extreme (bold
line of metric function and related Penrose diagram in the
right-down panel).} \label{Pen}
\end{figure}

The gauge potential that supports dynamical metric, from Eq.
(\ref{FE2}), follows as
\begin{equation}
h(r)=-\sqrt{\frac{d-2}{d-3}}\frac{q}{r^{d-3}}\; {}_{2}F_{1}\left( \left[ \frac{1}{2},\frac{d-3}{2(d-2)}\right] ,%
\left[ \frac{3(d-\frac{7}{3})}{2(d-2)}\right] ,-\Gamma \right) ,
\label{potential EM}
\end{equation}
where $\Gamma =\frac{(d-2)(d-3)q^{2}}{b ^{2}r^{2(d-2)}}$ and $q$
is integration constant which is related to the electric charge.
It is straightforward to show that the nonzero component of
electromagnetic field tensor is
$F_{tr}=\frac{\sqrt{(d-2)(d-3)}}{\sqrt{1+\Gamma }}
\frac{q}{r^{(d-2)}}$. In addition, regarding the nonzero
components of the gravitational field equation (Eq. (\ref{FE1})),
simultaneously, the metric function is obtained \cite{hendi3}
\begin{eqnarray}
f(r) &=&k-\frac{m_{0}}{r^{d-3}}+\left( \frac{4 b ^{2}-2\Lambda }{%
(d-1)(d-2)}\right) r^{2}+\frac{4(d-2)q^{2}}{(d-1)r^{2(d-3)}}\; {}_{2}F_{1}\left( \left[ \frac{1}{2},\frac{d-3}{2(d-2)}\right] ,%
\left[ \frac{3(d-\frac{7}{3})}{2(d-2)}\right] ,-\Gamma \right)
\nonumber \\
&& -\frac{4 b ^{2}r^{2}}{(d-1)(d-2)}\sqrt{1+\Gamma
}+\frac{m^{2}}{d-2}\left(
\sum\nolimits_{i=1}^{n}c^{i}c_{i}r^{2-i}\prod\nolimits_{j=2}^{i}(d-j)\right),
\label{metric function}
\end{eqnarray}
in which $m_{0}$ is an integration constant which is related to
the total mass of the black hole. Calculations confirm that there
is a curvature singularity at the origin which is covered by an
event horizon, and therefore, the solutions can be interpreted as
black holes \cite{hendi3}. It is also notable that by considering
different values for the parameters the roots of metric function
have different behaviors (see \cite{hendi3} for more details). In
order to study the conformal behavior of the solutions, one can
use the conformal compactification method to plot conformal
(Penrose) diagrams (see Fig. \ref{Pen} and also \cite{hendi3}).

\section{Action of the WDW patch}

According to the CA conjecture, the complexity of a boundary state
is proportional to the classical action of a region of spacetime
called the WDW patch which is the region in the bulk inclosed
between rays. The mentioned patch in the
adS-Einstein-Born-Infeld-massive black hole spacetime with
considering the case with two horizons (outer one is extreme) is
evolved in time from $t_{1}$ to $t_{2}$.

We focus on the rate change of action as a function of time rather
than its absolute value. The different parts that may contribute
to the action growth are bulk region $V_{1}$ and $V_{2}$, and
null-null surface joints $A$, $B$, $C$ and $D$ (for more details
see \cite{poisson})
\begin{eqnarray}
\partial S &=&\int_{V_{1}}dr\ dt\ d^{d-2}x_{i}\sqrt{-g}~\pounds %
-\int_{V_{2}}dr\ dt\ d^{d-2}x_{i}\sqrt{-g}~\pounds +\frac{1}{8\pi G}%
\int_{B}a_{B}\sqrt{\gamma }d^{d-2}x_{i}  \nonumber \\
&&-\frac{1}{8\pi G}\int_{A}a_{A}\sqrt{\gamma }d^{d-2}x_{i}+\frac{1}{8\pi G}%
\int_{D}a_{D}\sqrt{\gamma }d^{d-2}x_{i}-\frac{1}{8\pi G}\int_{C}a_{C}\sqrt{%
\gamma }d^{d-2}x_{i},  \label{action 2}
\end{eqnarray}
where $\gamma $ is the determinant of induced metric and the
integrant $a$ has the following form
\begin{equation*}
a=\ln (-\frac{1}{2}N.\bar{N}),
\end{equation*}
in which $N$ is the future-directed null normal to the left-moving
null surface and $\bar{N}$ denotes the future-directed null normal
to the right-moving null surface. For calculating the contribution
of volume, we can write
\begin{equation}
S_{V_{1}}-S_{V_{2}}=-\frac{V_{d-2}}{16\pi }\delta t\left[ 4\Lambda
r^{d-1}-m^{2}X+BI\right] _{r_{-}}^{r_{+}},  \label{bulk}
\end{equation}
where $BI$ and $X$ are
\begin{equation}
BI=-\frac{8 b ^{2}r^{d-1}}{(d-1)(d-2)}\left[1+\;{}_{2}F_{1}\left([-\frac{1}{2},\frac{1-d%
}{2d-4}],[1,\frac{1-d}{2d-4}],-\frac{(d-2)(d-3)q^{2}}{2 b
^{2}r^{2d-4}}\right) \right] \label{bi term}
\end{equation}
\begin{eqnarray}
&&X=\frac{r^{(d-5)}}{(d-1)(d-2)}%
(-48c_{4}c^{4}+2c_{1}r^{4}-4c_{2}r^{4}+12c_{3}r^{4}+48c_{4}c^{4}+2cc_{1}r^{3}d+2c_{2}c^{2}r^{2}d^{2}
\nonumber \\
&&-6c^{2}c_{2}r^{2}d-2cc_{2}r^{3}d^{2}+6cc_{2}r^{3}d+2c^{3}c_{3}rd^{3}-12c_{3}c^{3}rd^{2}+22c_{3}c^{3}rd-3c^{2}c_{3}r^{2}d^{3}+
\nonumber \\
&&9c_{3}c^{2}r^{2}d^{2}-6c_{3}c^{2}r^{2}d+6cc_{3}r^{3}d^{2}-18cc_{3}r^{3}d-4c_{4}c^{3}rd^{4}+32c_{4}c^{3}rd^{3}-92c_{4}c^{3}rd^{2}+
\nonumber \\
&&112c_{4}c^{3}rd+12c_{4}c^{2}r^{2}d^{3}-36c_{4}c^{2}r^{2}d^{2}+24c_{4}c^{2}r^{2}d-24cc_{4}r^{3}d^{2}+72cc_{4}r^{3}d-c_{1}r^{4}d-
\nonumber \\
&&4cc_{2}r^{3}+2c_{2}r^{4}d+12cc_{3}r^{3}-6c_{3}r^{4}d+2c_{4}c^{4}d^{4}-20c_{4}c^{4}d^{3}+70c_{4}c^{4}d^{2}-100c_{4}c^{4}d-
\nonumber \\
&&48c_{4}c^{3}r-48c_{4}cr^{3}+24c_{4}r^{4}d-2cc_{1}r^{3}-12c_{3}c^{3}r+4c_{2}c^{2}r^{2}).
\label{massive}
\end{eqnarray}

Before calculating the contributions for joints, we transform $N$
and $\bar{N}$ under an affine parametrization
\begin{equation*}
N_{\alpha }=-b_{1}\partial _{\alpha }(t-r^{\ast }),\ \ \ \ \ \
\bar{N} =b_{2}\partial _{\alpha }(t+r^{\ast }),
\end{equation*}
in which  $b_1$ and $b_2$ are two arbitrary positive constants and
$r^{\ast}$ is defined as
\begin{equation}
r^{\ast }=\int \frac{dr}{f(r)}. \label{r-star}
\end{equation}

Finally the contributions for joints are calculated as
\begin{equation}
S_{B}-S_{A}=\frac{V_{d-2}\delta t}{16\pi }\ \left[r^{d-2}\left( \frac{d-2}{r}%
f(r)\ln [\frac{f(r)}{b_{1}b_{2}}]+f'(r)\right)\right] _{r_{A}},
\label{joint1}
\end{equation}
\begin{equation}
S_{D}-S_{C}=\frac{V_{d-2}\delta t}{16\pi }\ \left[r^{d-2}\left( \frac{d-2}{r}%
f(r)\ln [\frac{f(r)}{b_{1}b_{2}}]+f'(r)\right)\right] _{r_{C}},
\label{joint2}
\end{equation}
where $ f'(r)=\frac{df(r)}{dr}$. Now, by combining the volume and
joint contributions, we obtain
\begin{equation}
\frac{dS}{dt}=-\frac{V_{d-2}}{16\pi }\left[ 4\Lambda
r^{d-1}-m^{2}X+BI\right] _{r_{-}}^{r_{+}}+\frac{V_{d-2}}{16\pi
}\left[r^{d-2}\left( \frac{d-2}{r}f(r)\ln [
\frac{f(r)}{b_{1}b_{2}}]+f'(r)\right)\right] _{r_{A}}^{r_{C}}.
\label{rate}
\end{equation}

At the late time $r_{A}$ and $r_{C}$ approach, respectively, to
$r_{-}$ and $r_{+}$ and $f(r)$ goes to zero, and therefore, Eq.
(\ref{rate}) becomes
\begin{equation}
\frac{dS}{dt}=-\frac{V_{d-2}}{16\pi }\left[ 4\Lambda
r^{d-1}-m^{2}X+BI\right] _{r_{-}}^{r_{+}}+\frac{V_{d-2}}{16\pi
}\left[r^{d-2}\left( f'(r)\right)\right] _{r_{-}}^{r_{+}}.
\label{rate2}
\end{equation}

Regarding the above relation and Eq. (\ref{C-A}) the rate of
complexity (the left hand side of the Eq. (\ref{bound}) ) will be
calculated. Here, we should determine the right hand side of Eq.
(\ref{bound}). According to the presence of nonlinear
electrodynamics and massive term in the action, it is notable that
the equation for upper bound of the rate of complexity must be
modified by additional terms as
\begin{equation}
\frac{\partial C}{\partial t} \leq \frac{2}{\pi \hbar }\left[
\left( M-\mu Q-B b -\sum\nolimits_{i=1}C_{i}c_{i}\right)_{r_+}
-\left( M-\mu Q-B b -\sum\nolimits_{i=1}C_{i}c_{i}\right)
_{r_-}\right]. \label{Llyod}
\end{equation}

Now, we regard the electric charge, massive parameters and the
nonlinearity parameter as extensive parameters and then we obtain
their intensive conjugates. The chemical potential is intensive
parameter which conjugates to the electric charge and is obtained
from integration of the electromagnetic tensor
\begin{equation}
\mu =-\frac{\sqrt{(d-2)(d-3)}q}{(d-3)r^{d-3}}\;{}_{2}F_{1}\left([\frac{1}{2},\frac{d-3%
}{2(d-2)}],[\frac{3(d-\frac{7}{3})}{2(d-2)}],-\frac{(d-2)(d-3)q^{2}}{b
^{2}r^{2d-4}}\right).
\end{equation}

Next, we should obtain conjugate quantities of the massive
parameters. For this purpose, the total mass should be obtained
from Hamiltonian approach with the following explicit form
\begin{equation}
M=\frac{(d-2)V_{d-2}}{2k^{2}}m_{0},
\end{equation}
where $m_{0}$ can be replaced from the fact that the metric
function vanishes at the horizon. Thus, we can write
\begin{eqnarray}
M &=&\frac{(d-2)V_{d-2}}{2k^{2}}r_{+}^{d-3}\left[k-\frac{4 b ^{2}r_{+}^{2}}{%
(d-1)(d-2)}\sqrt{1+\frac{q^{2}(d-2)(d-3)}{b ^{2}r_{+}^{2d-4}}}+r_{+}^{2}(%
\frac{\Lambda }{2(d-2)(d-3)}+ \right. \nonumber \\
&& \frac{4 b
^{2}}{(d-1)(d-2)})+\frac{4(d-2)q^{2}}{(d-1)r_{+}^{2d-6}}
\;{}_{2}F_{1}([\frac{1}{2},\frac{d-3}{2d-4}],[\frac{3d-7}{2d-4}],-\frac{%
q^{2}(d-2)(d-3)}{b ^{2}r_{+}^{2d-4}})+  \nonumber \\
&& \left.
m^{2}(\frac{cc_{1}r_{+}}{d-2}+c^{2}c_{2}+\frac{(d-3)c^{3}c_{3}}{r_{+}}+
\frac{(d-3)(d-4)c_{4}c^{4}}{r_{+}^{2}})\right].  \label{total
mass}
\end{eqnarray}

In order to calculate the conjugate quantities, we can use the
first law of thermodynamics. Differentiating of $M$ with respect
to the massive parameters gives their conjugates, as
\begin{equation}
C_{1}=\frac{dM}{dc_{1}}=\frac{ V_{d-2}}{2}m^{2}c k r_{+}^{d-2},
\end{equation}
\begin{equation}
C_{3}=\frac{dM}{dc_{3}}=\frac{(d-2)(d-3)m^{2}kV_{d-2}}{2}r_{+}^{d-4}c^{3},
\end{equation}
\begin{equation}
C_{4}=\frac{dM}{dc_{4}}=\frac{(d-2)(d-3)(d-4)m^{2}kV_{d-2}}{2}%
r_{+}^{d-5}c^{4}.
\end{equation}

%%%%%%%%%%%%%%%%%%%%%%%%%%%%%%%%%%%%%%%%%%%%%%%%%%%%%%%%%%%%%%%%%%%%%%%%%%

%%%%%%%%%%%%%%%%%%%%%%%%%%%%%%%%%%%%%%%%%%%%%%%%%%%%%%%%%%%%%%%%%%%%%%%%%%

As one can see, we did not calculate the conjugate of $c_{2}$. It
is due to the fact that this term has no contribution to the Smarr
relation, and consequently, has no contribution to Eq.
(\ref{Llyod}) (see \cite{hendi8} for more details). Regardless of
$c_{2}$, in general, there are $d$ massive parameters, and
therefore, $d$ conjugate quantities. The $n^{th}$ conjugate
quantity related to $c_{n}$ is introduced in the following form
\begin{equation}
C_{n}=\frac{m^{2}kV_{d-2}}{2}c^{n}r_{+}^{d-n-1}\prod_{i=2}^{n}(d-i).
\end{equation}

Eventually intensive parameter which is conjugate to the
nonlinearity parameter, $b$, is
\begin{eqnarray}
B &=&\frac{dM}{db }=\frac{(d-2)kV_{d-2}r_{+}^{d-1}}{2} \left[ \frac{8 b }{%
(d-1)(d-2)}\left(1-\sqrt{1+\frac{(d-2)(d-3)q^{2}}{b ^{2}r_{+}^{2d-4}}}\right)+ \right.\nonumber \\
&&\frac{4q^{2}(d-3)r_{+}^{4-2d}}{b (d-1)\sqrt{1+\frac{(d-2)(d-3)q^{2}}{%
b ^{2}r_{+}^{2d-4}}}}+\frac{4(d-2)^{2}(d-3)^{2}q^{4}}{(3d-7)(d-1)b
^{3}}r_{+}^{12-4d} \times \nonumber  \\
&&\left. \;{}_{2}F_{1}\left([\frac{3}{2},\frac{3d-7}{2d-4}],[\frac{5d-11}{2d-4}],-\frac{%
q^{2}(d-2)(d-3)}{b ^{2}r_{+}^{2d-4}}\right) \right]
\end{eqnarray}

At this situation, we calculate both the right and left hand sides
of Eq. (\ref{Llyod}), and therefore, we can numerically check the
validity of the inequality (\ref{Llyod}). We can plot a figure for
both LHS and RHS of Eq. (\ref{Llyod}) versus $c_{1}$, $c_{3}$, $m$
and $b$. According to these plots, we find that the inequality
(\ref{Llyod}) is always held {because the diagram for the RHS is
constant and the diagram for LHS is negative and with increasing
of each parameter decreases therefore $RHS-LHS$ has always
positive value.}

\section{Shockwave and discontinuity}

In this section, we study CA duality in the presence of a bulk
shockwave which is dual to the insertion of a perturbation in the
past of the thermofield double state. The motivation of adding
perturbation is to investigate entanglement between two states
before and after adding perturbation. Suppose that two typical
systems $A$ and $B$ (which are small subsystems in left and right
state) at time $t=0$ are highly entangled. Then we consider the
effect of injecting a small amount of energy $E$ into the left
system, by throwing a few quanta towards the horizon at time
$-t_w$. One expects that the CFTs dual to black holes have
sensitive dependence on the initial conditions, and this small
perturbation should touch off chaotic behavior in the left
\textbf{theory} and then should have less entanglement between $A$
and $B$ \cite{shenker}. The added complexity by this perturbation
can be understood in term of the minimal quantum circuit needed to
apply the Heisenberg operator $W(t_{w})=e^{iHt_{w}}We^{-iHt_{w}}$
to the thermofield double state. According to Ref.
\cite{susskindcsg,susskinddd}, we expect a partial cancelation of
the forward and backward time evolutions generating $W(t_{w})$,
and therefore, the total additional complexity for large $t_{w}$
is proportional to $2(t_{w}-t_{*})$ in which $t_{*}$ is the
scrambling time. For small values of $t_{w}$ the geometry will not
be substantially affected by the perturbation. The Schwarzschild
time evolution acts near the horizon as a boost.

%%%%%%%%%%%%%%%%%%%%%%%%%%%%%%%%%%%%%%%%%%%%%%%%%%%%%%%%%%%%%%%%%%%%%%%
%%%%%%%%%%%%%%%%%%%%%%%%%%%%%%%%%%%%%%%%%%%%%%%%%%%%%%%%%%%%%%%%%%%%%%%

For understanding the effect of shockwave, it is helpful to change
the coordinates $t$ and $r$ to $u$ and $v$ with the following
definition
\begin{equation*}
u=t+r^{\ast },\;\;\;   v=t-r^{\ast },
\end{equation*}
in which $r^{\ast}$ is defined in Eq. (\ref{r-star}). With
applying these change of coordinates, the line element changes to
\begin{equation}
ds^{2}=-f(r)dudv+r^{2}h_{ij}dx^{i}dx^{j},\ \ \ i,j=1,2,3,...,d-2.
\end{equation}

It is convenient to carry out the shockwave calculations in the
Kruskal-Szekeres coordinates. These coordinates can be defined
throughout the eternal black hole spacetime as
\begin{eqnarray*}
U =-e^{-\frac{2\pi }{\beta }u}, V=e^{\frac{2\pi }{\beta }
v} (right\;exterior\;region) \\
U =e^{-\frac{2\pi }{\beta }u}, V=e^{\frac{2\pi }{\beta }v} (black\;hole\;region) \\
U =e^{-\frac{2\pi }{\beta }u}, V=-e^{\frac{2\pi }{\beta }
v} (left\;exterior\;region) \\
U =-e^{-\frac{2\pi }{\beta }u}, V=-e^{\frac{2\pi }{\beta } v}
(white\;hole\;region)
\end{eqnarray*}
where $\beta$ is the inverse of temperature with the following
explicit form
\begin{equation}
\beta =\frac{4\pi }{\partial _{r}f(r)}|_{r=r_{+}}
\end{equation}

The null shell is injected from the left boundary at time
$t_{w}\rightarrow \infty $ with infinitesimal energy $\delta
\epsilon $. In addition, the stress energy distribution is highly
compressed in $u$ direction but stretched in $v$ direction, and we
can replaced it by a stress tensor that localized at $u=0$ horizon
\begin{equation}
T_{uu}=\frac{\delta \epsilon }{l^{d}}e^{\frac{2\pi t_{w}}{\beta
}}\delta (u).
\end{equation}

Since the shockwave makes a discontinuity in the metric at $u=0$,
it leads to a finite shift in $v$
\begin{equation}
\delta v=h\sim e^{\frac{2\pi (t_{w}-t_{\ast })}{\beta }}.
\end{equation}

%%%%%%%%%%%%%%%%%%%%%%%%%%%%%%%%%%%%%%%%%%%%%%%%%%%%%%%%%%%%%%%%%%%%%%%
%%%%%%%%%%%%%%%%%%%%%%%%%%%%%%%%%%%%%%%%%%%%%%%%%%%%%%%%%%%%%%%%%%%%%%%

Here, we are going to calculate the complexity in the geometry
which is perturbed by a spherically symmetric null shell falling
into the black hole. The null shell sets of a shockwave whose
physical manifestation is a null shift along the shockwave. The
metric is discontinuous along the $EG$, and so, if we select two
points with the same $r$ but different $t$, we should not expect
the action of both points to add up to zero. We calculate the
effect of discontinuity by comparing the two null surfaces
$u=\epsilon $ and $u=-\epsilon$ which approach to $EG$ when
$\epsilon \rightarrow 0$. The contribution of the discontinuity is
\begin{equation}
S_{discontinuity}=S_{E'G'}+S_{E''
G''}+S_{E'}+S_{G'}+S_{E''}+S_{G''}. \label{dis}
\end{equation}

The lines $E' G'$ and $E'' G''$ are null, and therefore, the first
two terms in Eq. (\ref{dis}) becomes zero. It is notable that the
metric function $f(r)$ is positive and negative for outside and
inside the horizon, respectively and therefore, the actions of
joints are
\begin{eqnarray*}
S_{E'}&=&\frac{\Omega ^{d-2}}{8\pi G}r^{d-2}\ln [f(r)], \\
S_{G'}&=&-\frac{\Omega ^{d-2}}{8\pi G}r^{d-2}\ln [-f(r)],
\\
S_{G''}&=&\frac{\Omega ^{d-2}}{8\pi G}r^{d-2}\ln [f(r)],
\\
S_{E''}&=&-\frac{\Omega ^{d-2}}{8\pi G}r^{d-2}\ln [-f(r)].
\end{eqnarray*}

Since the radius $r\rightarrow r_{h}$ for $\epsilon \rightarrow
0$, we expand the contributions of the joints $E'$, $G'$, $E''$
and $H''$ around $r_{h}$. The expansion of $r^{d-2}ln[-f(r)]$ (for
$r<r_h$) is
\begin{equation*}
r^{d-2}\ln [-f(r)]|_{r \rightarrow r_{h}}=\left(r_{h}^{d-2}\ln
[-(r-r_{h})H]\right)_{r<r_h},
\end{equation*}
where $H$ is a constant which is defined as
\begin{eqnarray*}
H &=&2k-\frac{8 b
^{2}r_{h}}{(d-1)}\sqrt{1+\frac{q^{2}(d-2)(d-3)}{b
^{2}r_{h}^{2(d-2)}}}+2r_{h}(d-2)\left(\frac{1}{l^{2}}+\frac{4 b ^{2}}{%
(d-1)(d-2)}\right) \\
&&2m^{2}(d-3)\frac{1}{r_{h}}\left[\frac{cc_{1}r_{h}}{d-2}+c_{2}c^{2}+\frac{%
(d-3)c^{3}c_{3}}{r_{h}}+\frac{(d-3)(d-4)c^{4}c_{4}}{r_{h}^{2}}\right]+ \\
&&\frac{4r_{h}(d-2)(d-3)q^{2}r_{h}^{-2(d-2)}}{(d-1)\sqrt{1+\frac{%
q^{2}(d-2)(d-3)}{b ^{2}r_{h}^{2(d-2)}}}}+\frac{4}{3}\frac{%
(d-2)^{3}(d-3)^{2}q^{4}}{(d-1)(d-\frac{7}{3})b ^{2}r_{h}^{2d-4}} \\
&&{}_{2}F_{1}\left([\frac{3}{2},\frac{d-3}{2(d-2)}+1],[\frac{3(d-\frac{7}{3})}{%
2(d-2)}+1],\frac{q^{2}(d-2)(d-3)}{b ^{2}r_{h}^{2(d-2)}}\right)+ \\
&&m^{2}\left[\frac{cc_{1}}{d-2}-\frac{(d-3)c^{3}c_{3}}{r_{h}^{2}}-\frac{%
2(d-3)(d-4)c^{4}c_{4}}{r_{h}^{3}}\right].
\end{eqnarray*}

On the other hand, by using the definition of Kruskal-Szekeres
coordinates, we can write
\begin{eqnarray}
UV &=&e^{-\frac{4\pi }{\beta
 }r^{\ast }}; for\;inside\;the\;horizon,  \nonumber \\
UV &=&-e^{-\frac{4\pi }{\beta }r^{\ast }};
for\;outside\;the\;horizon. \label{UV}
\end{eqnarray}

In addition, we can use the series expansion of $r^{\ast}$ for
$r\rightarrow r_{h}$ to obtain
\begin{equation*}
r^{\ast }|_{r\rightarrow r_{h}}=\frac{1}{H}\ln
((r-r_{h})H)_{r>r_{h}}.
\end{equation*}

Considering Eq. (\ref{UV}), one finds that $r^{\ast}$ can be
written as $r^{\ast }=-\frac{\beta }{4\pi}ln(\pm UV)$, and
therefore, it can be calculated at the points $E'$, $G'$, $E''$
and $G'$, as
\begin{eqnarray*}
r^{\ast }(E') &=&-\frac{\beta }{4\pi }\ln (\epsilon U_{0}^{-1}), \\
r^{\ast }(G') &=&-\frac{\beta }{4\pi }\ln (\epsilon V_{0}+\epsilon h), \\
r^{\ast }(E'') &=&-\frac{\beta }{4\pi }\ln (\epsilon U_{0}^{-1}+\epsilon h), \\
r^{\ast }(G'') &=&-\frac{\beta }{4\pi }\ln (\epsilon V_{0}),
\end{eqnarray*}
where $U_0$ and $V_0$ are two arbitrary positive constants. Now,
we are in a position to replaced our results in the action of
discontinuity, yielding
\begin{eqnarray*}
S_{E'}&=&-\frac{\beta H \Omega ^{d-2}}{32\pi ^{2}G}r_{h}^{d-2}\ln
(\epsilon U_{0}^{-1}), \\
S_{G'}&=&\frac{\beta H \Omega ^{d-2}}{32\pi ^{2}G}r_{h}^{d-2}\ln
(\epsilon V_{0}+\epsilon h), \\
S_{E''}&=&\frac{\beta H \Omega ^{d-2}}{32\pi ^{2}G}r_{h}^{d-2}\ln
(\epsilon U_{0}^{-1}\epsilon h),  \\
S_{G''}&=&-\frac{\beta H \Omega ^{d-2}}{32\pi ^{2}G}r_{h}^{d-2}\ln
(\epsilon V_{0}),
\end{eqnarray*}
where, in final form, we can obtain
\begin{equation}
S_{discontinuity}=\frac{\beta H \Omega ^{d-2}}{32\pi
^{2}G}r_{h}^{d-2}\ln ((1+V_{0}^{-1}h)(1+U_{0}h)).
\end{equation}

It is obvious that $h$, $U_0$ and $V_0$ are positive, and
therefore, $S_{discontinuity}$ is positive too and its value
depends only on $h$. For vanishing $h$, one finds that
$S_{discontinuity}$ vanishes too (since the effect of shockwave is
vanished) and for nonzero $h$, $S_{discontinuity}$ is an
increasing function of $h$. In addition, it is interesting to note
that for $d > 2$, increasing the horizon radius (or decreasing the
temperature) leads to increasing $S_{discontinuity}$. For
describing physical meaning of $S_{discontinuity}$ one should note
that at first complexity is calculated for the Einstein-Rosen (ER)
bridge. In this way a signal is sent to ER bridge and difficulties
which is faced to the exiting of the signal (definition of the
complexity) is proportional to the volume of ER bridge; Adding a
bulk of shockwave, in fact, distorts ER bridge and a discontinuity
is created which adds more difficulty (complexity) in exiting of
the signal.

\section{ Conclusion}

In this paper, we have examined the holographic complexity of the
black holes in Einstein-massive gravity in the presence of BI
electrodynamics with CA proposal. For this purpose, we have
calculated the complexity based on the action including of bulk
term and joints contributions. We have also obtained intensive
parameters based on the extended first law of thermodynamics and
their conjugate extensive quantities.

In addition, we have calculated the both side of Llyod equation,
separately, and found numerically that such inequality is always
held. Finally, we have regarded a geometrical perturbation in the
context of bulk shockwave and calculated the contribution of the
possible discontinuity. We have shown that the discontinuity is
affected by the horizon radius (and also temperature),
significantly.

According to the result of the rate of complexity for the case
Einstein-Born-Infeld-Massive black holes, we can compute some
limited cases as the subclass of our model:
Einstein-Maxwell-Massive black holes, Einstein-Born-Infeld black
holes and Einstein-Maxwell black holes which are well-known as the
Reissner-Nordstr\"{o}m black holes. The results are summarized as
follows: for the case:  Einstein-Maxwell-Massive black hole, we
should regard $b \longrightarrow \infty$ case with the following
result
\begin{eqnarray}
\frac{dC}{dt} &=&-\frac{V_{d-2}}{16\pi ^{2}h}\left[ 4\Lambda r^{d-1}-m^{2}X+%
\frac{4}{d-2}q^{2}r^{3-d}\right] _{r_{-}}^{r_{+}}+\frac{V_{d-2}}{16\pi ^{2}h}%
\lim_{b\rightarrow \infty }\left[ r^{d-2}f^{^{\prime }}(r)\right]
_{r_{-}}^{r_{+}} \nonumber \\
&\leq &\frac{2}{\pi h}[(M-\mu
Q-\sum\nolimits_{i=1}C_{i}c_{i})_{r_{+}}-(M-\mu
Q-\sum\nolimits_{i=1}C_{i}c_{i})_{r_{-}}].
\end{eqnarray}

For the case of the Einstein-BI black holes we can write:
\begin{eqnarray}
\frac{dC}{dt} &=&-\frac{V_{d-2}}{16\pi ^{2}h}\left[ 4\Lambda r^{d-1}+BI%
\right] _{r_{-}}^{r_{+}}+\frac{V_{d-2}}{16\pi ^{2}h}\left[
r^{d-2}f^{^{\prime }}(r)\right] _{r_{-}}^{r_{+}}|_{m=0} \nonumber  \\
&\leq &\frac{2}{\pi h}[(M-\mu Q-bB)_{r_{+}}-(M-\mu Q-bB)_{r_{-}}].
\end{eqnarray}

For the case of Reissner-Nordstr\"{o}m black holes the following
result is obtained
\begin{eqnarray}
\frac{dC}{dt} &=&-\frac{V_{d-2}}{16\pi ^{2}h}\left[ 4\Lambda r^{d-1}+\frac{4%
}{d-2}q^{2}r^{3-d}\right] _{r_{-}}^{r_{+}}+\frac{V_{d-2}}{16\pi ^{2}h}%
\lim_{b\rightarrow \infty }\left[ r^{d-2}f^{^{\prime }}(r)\right]
_{r_{-}}^{r_{+}}|_{m=0} \nonumber \\
&\leq &\frac{2}{\pi h}[(M-\mu Q)_{r_{+}}-(M-\mu Q)_{r_{-}}].
\end{eqnarray}

It is notable that these results are in agreement with the
previous works reported in Refs. \cite{cai2017,swingle2017}.

It is interesting to compute the complexity of different black
holes with more than two horizons in the presence of different
gauge fields and investigate their perturbation with the bulk of
shockwave, and then comparing their $S_{discontinuity}$ with each
other in order to find a possible relation between
$S_{discontinuity}$ and the gauge fields.

\end{document}